\documentclass[conference,letterpaper,10pt,onecolumn]{IEEEtran}

\usepackage{graphicx}
\usepackage{amsmath,amssymb}
\usepackage{cite}
\usepackage{hyperref}
\usepackage{booktabs}
\usepackage{xcolor}
\usepackage{microtype}
\usepackage{algorithm}
\usepackage{algorithmic}
\usepackage{array}
\usepackage{lipsum}
\usepackage{graphicx}
\usepackage{caption}
\usepackage{subcaption}
\usepackage{tcolorbox}
\usepackage{enumitem} 
\usepackage{cleveref} 
\usepackage{siunitx}  

\hypersetup{
    colorlinks=true,
    linkcolor=blue,
    filecolor=magenta,
    urlcolor=cyan,
    pdftitle={Human-Centered Shared Autonomy Frameworks},
    pdfauthor={MH Farhadi, Ali Rabiee, Sima Ghafoori, Anna Cetera, Reza Abiri},
    pdfsubject={Shared Autonomy},
    pdfkeywords={Shared Autonomy, Human-centered AI, Biosignal Processing},
    pdfstartview={FitH}
}

\IEEEoverridecommandlockouts

\makeatletter

\renewcommand{\maketitle}{%
  \begingroup
  \renewcommand{\thefootnote}{\fnsymbol{footnote}}%
  \def\@makefnmark{\hbox{\@textsuperscript{\normalfont\@thefnmark}}}%
  {\centering \@maketitle}%
  \@thanks
  \endgroup
  \setcounter{footnote}{0}%
  \let\maketitle\relax%
  \let\@maketitle\relax%
  \gdef\@thanks{}%
  \gdef\@author{}%
  \gdef\@title{}%
  \let\thanks\relax%
}
\def\@maketitle{%
  \newpage
  \null
  \vskip 2em%
  \begin{center}%
  \let\footnote\thanks
  {\Large\bfseries \@title \par}%
  \vskip 1.5em%
  {\normalsize
   \lineskip .5em%
   \begin{tabular}[t]{c}%
   \@author
   \end{tabular}\par}%
  \end{center}%
  \par
  \vskip 1.5em%
}
\makeatother

\let\oldabstract\abstract
\let\oldendabstract\endabstract
\renewenvironment{abstract}{\oldabstract\noindent}{\oldendabstract\vspace{1em}}

\begin{document}

\title{Human-Centered Shared Autonomy for Motor Planning, Learning, and Control Applications}

\author{
\IEEEauthorblockN{MH Farhadi\textsuperscript{1,*}, Ali Rabiee\textsuperscript{1,*}, Sima Ghafoori\textsuperscript{1,*}, Anna Cetera\textsuperscript{1,*}, Wei Xu\textsuperscript{2, 3}, and Reza Abiri\textsuperscript{1, 3}}
\IEEEauthorblockA{\textsuperscript{1}Translational Neurorobotics Laboratory\\ Department of Electrical, Computer, and Biomedical Engineering\\
University of Rhode Island\\
Kingston, RI, USA}
\IEEEauthorblockA{\textsuperscript{2}Human-Centered AI (HCAI) Labs\\
Los Angles, CA, USA}
\IEEEauthorblockA{\textsuperscript{3}Senior Authors}
\IEEEauthorblockA{\textsuperscript{*}Equal Contributions}
}

\maketitle

\begin{abstract}
With recent advancements in AI and computation tools, intelligent paradigms have emerged to empower different fields such as shared autonomy and human-machine teaming in healthcare with new capabilities. Advanced AI algorithms (e.g., reinforcement learning) can be trained and developed to autonomously make individual decisions to achieve desired plan and motion goals. However, such independent decisions and goal achievements might not be ideal within healthcare, where human intent plays a pivotal and crucial role in guiding human-machine paradigms. This chapter presents a comprehensive review of human-centered shared autonomy AI frameworks, particularly by focusing on upper limb biosignal-based machine interfaces and their associated end-effector based motor control systems, including computer cursors, robotic arms, and planar robotic platforms. Our primary focus is delving into motor planning, learning (rehabilitation), and controls, specifically covering conceptual foundations of human-machine teaming in reach-and-grasp tasks, analyzing both the theoretical principles and practical implementations from both human and machine perspectives. We extended the discussion in each section to elaborate on how human and machine inputs can be blended as shared autonomy paradigms, with the healthcare applications. The chapter examines topics on human factors, biosignal processing for intent detection, shared autonomy approaches in brain-computer interfaces (BCI), rehabilitation, robots, assistive robotics, and finally, Large Language Models (LLM) as the next frontier. With a foundation based on human-centered factors, we also proposed adaptive shared autonomy AI as a potential high-performance paradigm for the two interactive human and AI agents. We identified current challenges in human-centered shared autonomy implementation and proposed future directions, particularly examining the roles of AI reasoning agents in advancing these systems. Through this analysis, we bridge neuroscientific understanding with robotics approaches to develop more intuitive, effective, and ethical human-machine teaming frameworks. 

\end{abstract}


\vspace{2em}
\tableofcontents
\vspace{1em}

\section{Introduction}

The symbiotic integration of human capabilities with intelligent machines is rapidly transforming our approach to complex motor tasks, particularly in domains requiring nuanced interaction and assistance, such as healthcare and robotics \cite{Ajoudani2018, abiri2024toward}. Within this evolving landscape, human-centered shared autonomy has emerged as a pivotal paradigm. This chapter embarks on a comprehensive exploration of shared autonomy (traditionally known as shared control) frameworks. Our focus is primarily on end-effector based systems, encompassing a range of applications from intuitive computer cursor control and sophisticated robotic arm manipulation to advanced rehabilitation and assistive robots within the healthcare sector. The core tenet of these systems lies in creating a collaborative partnership where the human user and the semi-autonomous system work together by leveraging their respective strengths to achieve a common goal \cite{Losey2018}.

The imperative for human-centered approaches stems from the need to ensure that autonomous systems remain aligned with user intentions, adapt to individual user needs and capabilities, and maximize tasks performance all while maintaining a feeling of agency and trust in users \cite{collier2025sense}. This is particularly critical when dealing with biosignals (such as electromyography (EMG) or electroencephalography (EEG) ) as the primary communication channel between the human and the machine \cite{Wolpaw2002}. These signals, while rich in information about user intent, are often complex, noisy, and highly variable, posing significant challenges for robust interpretation and seamless translation into machine actions \cite{Farina2010}. Shared autonomy addresses these challenges by not seeking to fully automate tasks, nor to place the entire burden of control on the human, but rather to intelligently arbitrate and blend control inputs from both ends \cite{Ajoudani2018}.

This chapter will delve into the conceptual foundations of human-machine teaming, examining how principles from cognitive science, neuroscience, and robotics converge to inform the design of effective shared control systems. We will analyze both the theoretical underpinnings and the practical implementations of these frameworks, tracing their evolution and impact. A significant portion of our discussion will be dedicated to the critical role of biosignal processing in accurately detecting human intentions \cite{He2020}. This involves exploring advanced algorithms for signal acquisition, feature extraction, and pattern recognition that enable the machine to infer the user's desired actions and goals.

We will survey the diverse applications of shared autonomy across various domains. In Brain-Computer Interfaces (BCIs), shared control offers a pathway to more intuitive and reliable operation of assistive devices for individuals with severe motor impairments \cite{millan2010combining}. Shared autonomy facilitates personalized therapy by dynamically adjusting the level of assistance based on the patient's engagement and performance, promoting motor learning and recovery \cite{MarchalCrespo2009}, ultimately empowering users by enhancing their ability to perform activities of daily living with greater independence and precision.

The burgeoning field of Large Language Models (LLMs) is also beginning to intersect with shared autonomy, offering new avenues for more naturalistic interaction, high-level task understanding, and even AI-driven reasoning to support this collaborative process \cite{Brohan2023}.

Despite significant advancements, the implementation of truly human-centered shared autonomy faces ongoing challenges. These include ensuring seamless and intuitive interaction, developing robust co-adaptation mechanisms, addressing ethical considerations related to user privacy, autonomy, and potential for algorithmic bias \cite{FoschVillaronga2021}, and validating system efficacy in real-world scenarios \cite{Hayes2013}. This chapter will identify these critical challenges and propose directions for development.

Our goal is to provide researchers, practitioners, and students with a thorough understanding of the principles, technologies, applications, and future horizons of human-centered shared autonomy. This chapter provides a unified analysis of shared autonomy for motor control by connecting three distinct applications: Brain-Computer Interfaces (BCIs), rehabilitation, and assistive robotics \cite{abiri2024toward}. Our contribution is to establish a common technical framework for these historically separate fields, one grounded in the mapping of user biosignals to control policies via adaptive arbitration \cite{He2020, jain2015assistive}. By examining methods—from Bayesian inference to modern deep learning architectures—through this unified lens, we identify shared challenges and opportunities for creating models that generalize across these human-in-the-loop systems \cite{lotte2022coadaptive, rabiee2025learning}.

\section{Human-centered Shared Autonomy Principles}

The effective integration of intelligent systems into human endeavors necessitates a paradigm shift from automation-centric designs towards those that prioritize a robust and adaptive synergy with the human user. Human-centered shared autonomy (HCSA) is at the forefront of this change, representing a field dedicated to the development of systems wherein human and artificial intelligence establish a cohesive and symbiotic partnership. The principles underpinning HCSA are not merely design heuristics; they constitute a foundational philosophy critical for realizing the transformative potential of autonomous technologies, particularly in domains with significant human consequence, such as healthcare, assistive robotics, and complex operational environments. Adherence to these principles is paramount for the creation of systems that are not only technically proficient but also intuitive, trustworthy, responsive, and ultimately, empowering for the individuals they are intended to assist \cite{Ajoudani2018}.

This section distills the core design principles of human‑centered shared autonomy (HCSA), detailing (i) why the human remains team‑leader, (ii) how assistance arbitration paradigms govern control blending, and (iii) how intent inference closes the loop.

\subsection{Human as the leader in human-AI teams}

A central principle within numerous shared‑autonomy paradigms, particularly those involving critical decision‑making or operation in unpredictable environments, is placing the human as the leader within this human‑AI teamwork \cite{Losey2018, Nikolaidis2013SMM}. This principle acknowledges that while artificial intelligence can display remarkable proficiency in perception and execution, autonomous systems often lack the generalized intelligence, common‑sense reasoning, and nuanced comprehension of human values needed to navigate real‑world complexities. Accordingly, the human’s leadership typically involves defining high‑level objectives, making strategic decisions at critical junctures, and actively monitoring the system’s performance and behavioral patterns—rather than issuing continuous low‑level control commands \cite{Beer2014LORA, Madduri2023, Ajoudani2018}. Positioning the human in this capacity ensures the autonomous system functions as a capable collaborator whose actions remain congruent with the human’s objectives and ethical precepts \cite{Cummings2004Accountability}; it also preserves accountability and the user’s sense of agency —the experience of volition and control over one’s actions and outcomes \cite{collier2025sense}—while enabling timely intervention to rectify or override system behavior when it deviates from expectations or when unforeseen circumstances arise \cite{Sheridan2005HFAI}.

\begin{figure}[!h]            
  \centering                    
  \includegraphics[width=0.8\linewidth]{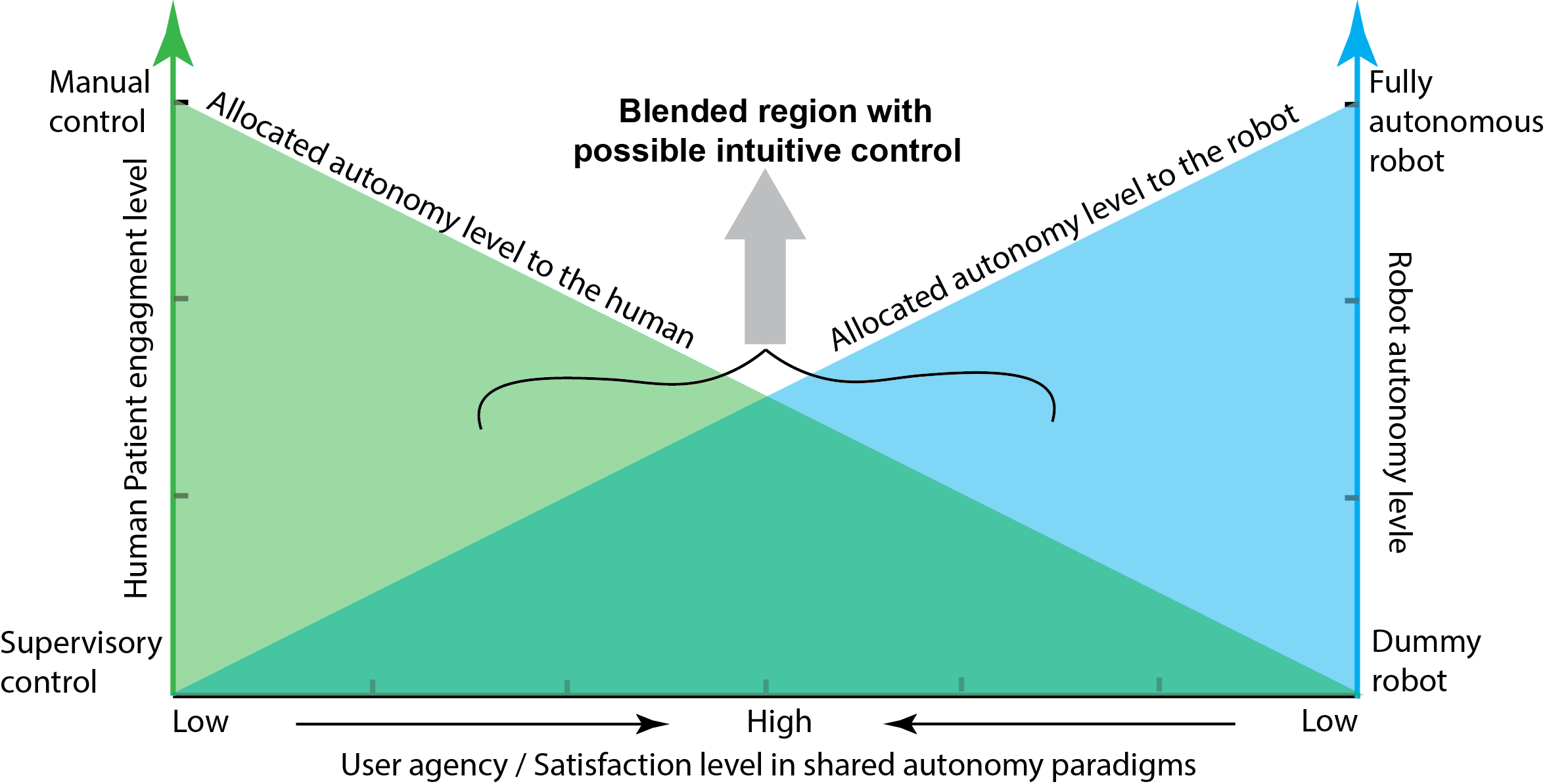}
  \caption{Conceptual autonomy–agency continuum for shared autonomy paradigms.}
  \label{fig:user_agency}         
\end{figure}

To facilitate this, systems frequently incorporate adjustable autonomy (also termed adaptive automation \cite{Li2015RoleAdapt}). This functionality allows the level of AI assistance to be dynamically modulated, either via explicit user directives, context-aware algorithms that adapt to shifting task demands, or implicitly through the inference of the user's cognitive state (e.g., fatigue, engagement levels). The user must be able to formulate an accurate mental model of the system's capabilities, its current operational status, its decision-making rationale (to an appropriate degree of abstraction), and its probable future actions. Simultaneously, the system adapts to the user's learning curve and evolving understanding. This understanding empowers the user to supervise effectively and anticipate the necessity for intervention.

\subsection{Interaction paradigms and levels of automation}

\begin{table}[t]
\centering
\caption{Overview of Arbitration Policies, Evaluation Environments, and Functionalities}
\label{tab:arbitration_policies_expanded}
\begin{tabular}{@{}p{4.5cm}p{5.5cm}p{6cm}@{}} 
\toprule
\textbf{Reference} & \textbf{Arbitration Policy} & \textbf{Evaluation Environment \& Functionality Tested} \\
\midrule
Losey et al. \cite{Losey2018} & Review: Co-activity, master-slave, teacher-student, collaboration; Arbitration definition. & Conceptual review; Case studies: Upper-limb robotic rehabilitation, haptic feedback for prosthetics. \\
\addlinespace
Franceschi et al. \cite{Franceschi2023TASE} & Role Arbitration (RA) via Game Theory (Coop. \& Non-Coop.); Dynamic $\alpha$ (blending) via Fuzzy Logic or force. & pHRI with UR5 robot; Co-transportation/manipulation: obstacle avoidance, via-point reaching. \\
\addlinespace
Madduri et al. \cite{Madduri2023} & Review: Co-adaptation in User-Machine Interfaces (UMI); Leader-follower; Game theory for user-decoder. & Conceptual review: Co-adaptive motor control interfaces (BCI, myoelectric, kinematic). \\
\addlinespace
McFarland et al. \cite{McFarland2008} & BCI control mapping (EEG to cursor). (Level of automation in BCI input translation). & Noninvasive BCI; 2D cursor movement \& target selection on screen; Emulates mouse operation. \\
\addlinespace
Sarasola-Sanz et al. \cite{SarasolaSanz2024} & Hybrid BMMI; Blended control (EEG-triggered EMG + robot assist). & 7-DoF Upper limb exoskeleton; Stroke rehabilitation: functional tasks (grasping, pointing). \\
\addlinespace
Jain and Argall \cite{JainArgall2019} & Probabilistic intent recognition; Blended shared control based on intent confidence. & 6-DoF MICO robotic arm; Assistive robotics: object retrieval/manipulation with joystick \& Sip-N-Puff. \\
\addlinespace
Franceschi et al. \cite{Franceschi2022ICRA} & RA (Cooperative Game Theory); Dynamic $\alpha$ via interaction force. & UR5 robot for pHRI; Planar circular trajectory following with human intervention. \\

\addlinespace
Nikolaidis and Shah \cite{Nikolaidis2013SMM} & Shared Mental Models for H-R Teaming; Robot adapts plan based on human's state/actions. (Arbitration via plan adaptation). & Simulated and Baxter robot; Physical assembly tasks (IKEA furniture). \\
\addlinespace
Ajoudani et al. \cite{Ajoudani2018SharedControl} & Shared control in HRI; Control sharing levels (human/robot lead, cooperation). & Broad review: pHRI, industrial co-manipulation, assistive robotics. \\
\addlinespace
Li et al. \cite{Li2015RoleAdapt} & Continuous Role Adaptation via Game Theory and Policy Iteration; Variable impedance based on Nash equilibrium. & pHRI experimental setup; Cooperative manipulation tasks (trajectory following). \\
\addlinespace
Mörtl et al. \cite{Mortl2012RoleOfRoles} & Leader-Follower role allocation analysis in pHRI; Energy exchange based role identification. & 1-DoF haptic knob, virtual crank, Dyadic co-manipulation (human-human and human-robot). \\

\bottomrule
\end{tabular}
\end{table}

The level of automation in shared autonomy emerges from the combination of interaction paradigm and arbitration scheme. An interaction paradigm describes the fundamental roles and collaborative structure within the human-robot team, as outlined in several works \cite{Losey2018, Jarrasse2012Framework}. To implement these different paradigms, we need different arbitration schemes 
for blending the input of each party. Table~\ref{tab:arbitration_policies_expanded} 
provides a comprehensive overview of arbitration policies across various shared 
autonomy implementations, highlighting the diversity of approaches in current research. In co-activity, the human and robot work within a shared environment, often on independent tasks that contribute to a broad objective, requiring spatial and temporal coordination. The master-slave paradigm places one agent, typically the human, in a position to dictate actions or trajectories, which the other agent (the robot) executes or supports—a common mode in teleoperation applications. In the teacher-student paradigm, one agent guides or adjusts the other to promote learning or rehabilitation; for example, a robot therapist might adaptively guide a patient's movements. Finally, collaboration in a partner-partner sense implies that human and robot act as peers, dynamically sharing responsibilities and control to achieve tasks that neither could perform as effectively in isolation, often involving mutual adaptation.

To implement these different paradigms, we need different arbitration schemes for blending the input of each party. Arbitration schemes specify the mechanisms for allocating or sharing control during task execution, determining "who controls what and when". These schemes range from fixed autonomy, where control authority is assigned a priori and remains static within task segments, to mode-switching, where control authority is explicitly transferred between agents at discrete junctions such as clear phase boundaries or specific triggers. Parallel schemes allow the human and robot to act simultaneously but on different, separable aspects of the task, often corresponding to disjoint Degrees of Freedom—for instance, the human might manage the positioning of an end-effector while the robot handles its orientation or grasp force. At the most fluid end of the spectrum, continuous policy-blending combines inputs or control policies from both the human ($u_{human}$) and the robot ($u_{AI}$) continuously to generate a single command $u_{final}$ for the system.
A common mathematical formulation for this is a weighted sum \cite{dragan2013policy, jain2015assistive}:
\begin{equation}
u_{final} = (1 - \alpha)u_{human} + \alpha u_{AI}
\end{equation}

The blending coefficient, $\alpha \in [0, 1]$, dictates the relative contribution of each agent, with $\alpha = 0$ signifying full human control and $\alpha = 1$ full robot control. Intermediate values represent varying degrees of shared control. The specific terms being blended ($u_{human}$, $u_{AI}$) can represent various control aspects such as forces, velocities, or position commands, depending on the application (e.g., as in \cite{SarasolaSanz2024} where control velocities from EMG and robot assistance are blended).

In continuous policy-blending, the coefficient $\alpha$ can be static or, more dynamically in HCSA, adaptive. An adaptive $\alpha$ allows the system to modulate the level of robot autonomy based on real-time factors such as confidence in intent inference \cite{jain2015assistive} or safety metrics like proximity to obstacles \cite{Franceschi2022ICRA, Franceschi2023TASE}. Game-theoretic arbitration \cite{Franceschi2022ICRA, Franceschi2023TASE, Li2015RoleAdapt} models the human and robot as players with distinct cost functions, determining dynamic arbitration strategies through evaluation of safety indices and task progress metrics.

The choice of interaction paradigm, arbitration scheme, and resultant level of automation profoundly influences the human user's experience. Systems that are opaque in their decision-making, unpredictable in their actions, or those that frequently override user input without clear justification can escalate frustration and erode user trust \cite{Hayes2013}. This directly impacts user agency and cognitive workload \cite{zhang2020survey, chen2014supervisory}. Optimal shared autonomy systems balance performance gains with preservation of user empowerment, often requiring adaptive arbitration that responds to individual user capabilities and preferences \cite{Beer2014LORA}.

\subsection{Decoding Human Intent for Adaptive Assistance}

Intent inference, where an autonomous system determines a human user's objectives or actions from limited observations, is fundamental to shared autonomy as it enables systems to anticipate user needs and provide timely assistance without explicit commands \cite{abiri2024toward}.

Three primary methodologies address intent inference in shared autonomy systems. Probabilistic methods explicitly model uncertainty in human behavior and noisy sensory data, maintaining probability distributions over possible user goals that update with new observations, as exemplified by recursive Bayesian filtering and POMDP frameworks \cite{jain2019probabilistic}. These approaches excel in interpretability but depend heavily on the accuracy of their models and priors. Data-driven methods leverage deep learning architectures like CNNs and RNNs to learn complex mappings from behavioral cues (EMG, gaze, kinematics) to user intentions, achieving high accuracy when sufficient training data is available \cite{mdpi_har_review_2024}. However, they often lack transparency and struggle with novel intents outside their training distribution. Model-based strategies employ explicit representations of human cognition, biomechanics, or task structures to interpret user behavior, such as inverse reinforcement learning that infers goals from observed actions \cite{annualreview_intent_inference_2024}. While highly interpretable, their performance is constrained by model fidelity.

The primary challenge in intent inference lies in extracting reliable information from inherently noisy, low-bandwidth biosignals that exhibit substantial inter- and intra-user variability \cite{frontiers_eeg_processing_review_2023}. The solution involves multimodal sensor fusion—combining complementary signals like EMG, EEG, and kinematics—coupled with personalization loops that adapt to individual users through online learning and calibration \cite{multimodal_emg_vision_grasp_2025, atan2024assistive}.

Intent inference serves as the critical link between the human-centered principles and practical implementation of shared autonomy systems. It directly influences how arbitration schemes adjust the blending coefficient $\alpha$ in real-time, with higher confidence in intent inference typically leading to increased autonomous assistance \cite{jain2015assistive}. The quality of intent inference fundamentally determines whether a shared autonomy system enhances or hinders the user's sense of agency. Accurate inference enables seamless assistance that feels like an extension of the user's will, while poor inference leads to frustrating conflicts between human and machine intentions \cite{gopinath2020active}. This challenge becomes particularly acute in biosignal-based systems, and robotic applications where neural and behavioral signals provide the primary communication channel, as we will explore in section \ref{bci}.

\subsection{Towards human-AI joint cognitive systems}

The trajectory of shared autonomy is pointing towards a paradigm that is more deeply integrated with human-AI teaming. The ultimate goal is the creation of a human-AI joint cognitive system, in which the human and machine function as a single, cohesive cognitive unit to achieve goals that are intractable for either partner alone \cite{Hollnagel2005}. This represents a shift in focus from purely optimizing physical task execution to improving the system's cognitive ergonomics, where the interactions are explicitly designed to minimize the user's cognitive load and mental fatigue \cite{Hancock2021}. The aim is to create a fluid, low-effort partnership. Rather than the user constantly directing the AI, the system begins to proactively assist based on task context and a predictive understanding of the user's high level goals \cite{Hoffman2019, Chen2025}.

Achieving this requires moving beyond reactive intent inference. For instance, instead of only decoding a user's immediate motor command, the system can learn to anticipate the next likely sub-task in a sequence (e.g., after picking up a cup, the next goal is likely bringing it to the mouth). This is a core challenge being explored in human-in-the-loop reinforcement learning, where the AI can be rewarded for actions that not only advance the task but also minimize the need for user corrections, thereby reducing cognitive burden \cite{Reddy2018}. Future biosignal-based interfaces will play a critical role, not by attempting to "read the user's mind," but by providing a real-time estimate of cognitive states like workload or surprise, which can serve as a feedback signal to modulate the AI's level of autonomy \cite{Muhl2014}.

The practical path towards this tighter integration also depends on improving the AI's transparency. The development of more explainable AI for robotics is crucial, as it allows the system to signal its intentions to the user, making its behavior more predictable and trustworthy \cite{Anjomshoae2019}. This creates a positive feedback loop: a more predictable system is easier to work with, which lowers cognitive load and builds the trust necessary for truly symbiotic human-AI partnership.

Having established the foundational principles of human-centered shared autonomy, 
we now examine their application across three critical domains: brain-computer 
interfaces, rehabilitation, and 
assistive technologies.

\section{Shared Autonomy in Brain-Computer Interface Domain}
\label{bci}
\subsection{Overview and gaps}
\begin{figure}[htbp]
    \centering
    \includegraphics[width=0.9\textwidth]{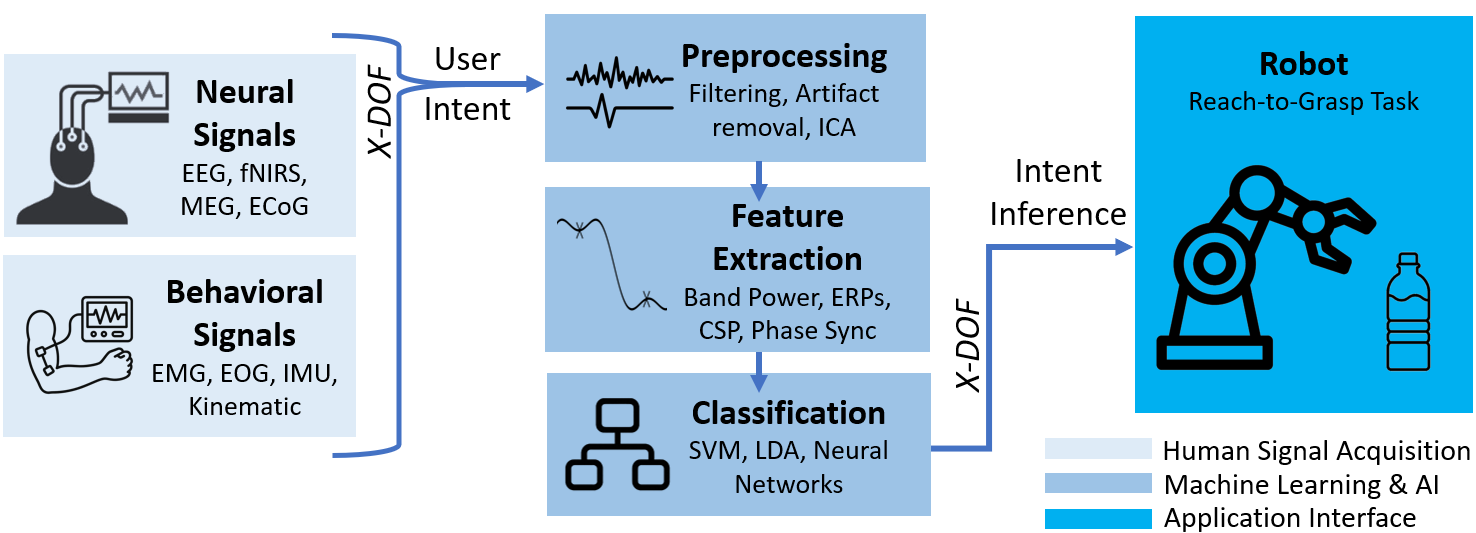}
    \caption{Traditional BCI pipeline for a reach-to-grasp task, illustrating intent decoding from neural and behavioral signals through preprocessing, feature extraction, and classification. Traditional BCIs operate under a sequential paradigm, preserving user input (X degrees of freedom, DOF).}
    \label{fig:bci_pipeline}
\end{figure}

Brain-Computer Interfaces (BCIs) have evolved from direct, low-level control systems toward more autonomous architectures that aim to reduce cognitive load, improve reliability, and enhance user experience \cite{abiri2019comprehensive}. Shared autonomy in BCIs involves dynamically blending user intention with system intelligence—commonly using biosignal-driven cues (e.g., EEG, EMG) and decision-making algorithms to assist users in completing motor or communication tasks.
Current approaches fall into three categories: Direct control involves users providing continuous or discrete commands (e.g., left/right movement) through modulation of specific brain rhythms or event-related potentials. While this offers high user agency, it is mentally demanding and slow, especially for multidimensional tasks \cite{Wolpaw2002, abiri2020usability}. Goal selection systems infer the user's intended goal from high-level signals and autonomously execute the corresponding action (e.g., selecting a target object). This reduces cognitive burden but requires accurate intent recognition and is often limited by low information transfer rates \cite{millan2010combining, kilmarx2018sequence}. Predictive shared control systems combine user input with predictive models (e.g., probabilistic inference, deep learning, or reinforcement learning) to collaboratively plan and execute actions. These systems can resolve ambiguity and correct errors, improving safety and efficiency \cite{ghasemi2022shared}.
However, these systems face key limitations: Signal variability presents challenges as EEG signals are highly non-stationary and subject to intra- and inter-subject variability, leading to unreliable decoding \cite{lotte2018review}. Calibration and adaptation issues arise because many systems require long calibration periods and fail to adapt dynamically to changes in neural patterns, fatigue, or user intent over time \cite{mcfarland2017eeg}. Trust and transparency become problematic as autonomy increases, since users may struggle to understand or trust the system's decisions, especially if they lack feedback or explanation mechanisms \cite{chen2014supervisory}. Cognitive load remains a concern because direct control BCIs impose significant mental effort, while poorly calibrated shared control models can increase user frustration due to erratic or incorrect assistance \cite{zhao2023human}. Lack of standardization in evaluation and benchmarking of shared autonomy approaches creates inconsistency, making it difficult to compare methods across studies.
Additionally, most systems rely on static control policies rather than adaptive models that learn and evolve with the user (Figure~\ref{fig:bci_pipeline}). These constraints highlight the need for co-adaptive, user-in-the-loop models that continuously learn from user behavior and feedback in BCI applications \cite {silversmith2021plug, natraj2024sampling}.

\subsection{Neural and behavioral signals}
Biosignal processing in shared autonomy systems combines neural and behavioral signals to infer user intention, monitor cognitive state, and support flexible human-robot interaction. These signals provide direct access to brain activity, muscle dynamics, and movement behavior, which together improve the robustness and reliability of human-in-the-loop systems.

Neural signals commonly used in non-invasive BCIs include electroencephalography (EEG), functional near-infrared spectroscopy (fNIRS), magnetoencephalography (MEG), and electrocorticography (ECoG). EEG is most commonly used because of its affordability, high temporal resolution, and portability \cite{lotte2018review, mcfarland2017eeg}. EEG measures brain activity by recording electrical signals from neurons in the cortex and researchers analyze EEG signals across different frequency ranges (delta, theta, alpha, beta, gamma) associated with different cognitive or motor states. It can also capture event-related potentials (ERPs), such as the P300 or N200, which indicate how the brain processes information and makes decisions \cite{pfurtscheller1999event}. MEG, while offering better spatial resolution than EEG, is costly and non-portable. ECoG, recorded directly from the cortical surface, provides high-fidelity signals but is invasive and thus limited to clinical contexts.

Behavioral signals provide physical expressions of brain activity and help decode user intention in shared autonomy. Electromyography (EMG) records electrical activity from skeletal muscles and is used to detect residual motor function or anticipate motion intention in prosthetics and exoskeletons \cite{clancy2002processing}. Electrooculography (EOG) captures eye movement potentials and is commonly applied to gaze-based interfaces, providing directional cues or confirmation signals. Inertial measurement units (IMUs), motion capture systems, and pressure sensors record kinematic data—such as joint angles, limb acceleration, or grip force—that reflect motor intent and biomechanics. These modalities can be used alone or combined with neural signals to build hybrid control systems \cite{schalk2004bci2000, muller2015combining}.

Multimodal integration of neural and behavioral signals strengthens shared autonomy by improving signal redundancy and reducing ambiguity in intention decoding. For instance, combining EEG with EMG can distinguish between imagined and attempted movements \cite{do2013brain, tortora2020hybrid} , while combining EOG with EEG allows gaze-contingent BCI control \cite{zander2010towards, allison2010bci}. However, multimodal systems also create problems with synchronization, computational complexity, and user burden. Creating interfaces that combine these signals effectively while maintaining speed and comfort remains a challenge for developing better BCI systems. Key areas for future work include adaptive feature fusion methods, wearable multi-sensor integration, and intelligent signal prioritization to unlock the capabilities of hybrid biosignal frameworks in autonomous BCI systems.

\subsection{Signal processing and feature extraction}

Transforming raw biosignals into actionable information requires a series of preprocessing, feature extraction, and classification steps. Preprocessing techniques such as bandpass filtering, artifact rejection (e.g., ICA for ocular and muscle artifacts), and baseline normalization are fundamental to improving signal quality and minimizing noise \cite{delorme2004eeglab, islam2016methods}. High-quality preprocessing is especially important in shared autonomy systems, where unreliable input can lead to incorrect system responses, ultimately reducing user trust and shared task success.
Feature extraction techniques identify relevant information from the preprocessed signals for downstream classification or regression tasks. In EEG, common features include band power (e.g., delta, theta, alpha, beta), event-related potentials (ERPs), phase synchrony \cite{lachaux1999measuring}, and spatial filters such as Common Spatial Patterns (CSP) or Filter Bank CSP (FBCSP) \cite{blankertz2008optimizing, ang2012filter}. For behavioral signals such as EMG or kinematic data, time-domain features (e.g., root mean square, variance) \cite{hudgins1993new} and frequency-domain features (e.g., FFT, wavelet transforms) \cite{englehart1999classification} are frequently used \cite{phinyomark2012feature}.

Recent developments in deep learning allows for end-to-end feature learning from raw biosignals using architectures such as convolutional neural networks (CNNs), recurrent neural networks (RNNs), and attention-based models \cite{lawhern2018eegnet, roy2019chrononet}. These approaches can automatically capture spatiotemporal dependencies across channels and time \cite{schirrmeister2017deep}, eliminating the need for manually computed features. However, deep learning models present challenges, particularly in shared autonomy scenarios where transparency, user interpretability, and real-time responsiveness are crucial. Black-box models may yield high accuracy but can be difficult to trust or troubleshoot in human-in-the-loop systems.

Additionally, the accuracy and performance of feature extraction pipelines is often user-dependent. Variability in neural and muscular signals between users—and even within the same user over time\cite{shenoy2006towards, vidaurre2011towards} —can reduce generalizability, leading to performance degradation without adaptive learning mechanisms. This is especially problematic in shared autonomy systems, where continuous co-adaptation between the user and machine is necessary to maintain harmonious user-system integration \cite{perdikis2018brain}. Systems that rely on static models or poorly calibrated features may either fail to assist the user effectively or override user input inappropriately, disrupting the collaborative balance.
To address these challenges, future feature extraction approaches in shared autonomy should aim for reliability to inter- and intra-user variability, interpretability and feedback integration for user understanding, low-latency and computational efficiency for real-time use, and adaptive and co-learning frameworks that evolve with the user.

\subsection{Classification and real-time intent inference}
Real-time intent decoding forms the backbone of shared autonomy brain-computer interface (BCI) systems, where human neural signals must be continuously translated into actionable commands while maintaining seamless human-machine collaboration. In these systems, the BCI component captures and decodes user intent from neural signals, while the autonomous component provides intelligent assistance to compensate for decoding uncertainties and reduce cognitive load. This collaborative framework requires robust real-time classification that can operate under the demanding constraints of continuous data streaming, minimal latency, and adaptive performance.

The process of real-time intent inference begins with continuous EEG signal acquisition, followed by rapid preprocessing, feature extraction, and classification within tight temporal windows—typically requiring complete processing cycles of 100-200 milliseconds to maintain natural interaction flow \cite{mason2007real, scherer2008toward}. However, this pipeline faces significant technical challenges that are particularly pronounced in shared autonomy applications. EEG signals suffer from inherently low signal-to-noise ratios \cite{nunez2006electric, cohen2014analyzing}, with neural activity of interest often buried beneath artifacts from muscle movement, eye blinks, and environmental interference \cite{lotte2018review, rezeika2018brain}. Motion artifacts become especially problematic in assistive applications where users may have involuntary movements or tremors. Additionally, maintaining sub-200ms latency is essential for preserving user trust and task performance, as delays in decoding disrupt the natural flow of shared control and can lead to user frustration or system abandonment \cite{zhao2023human}. These systems must also contend with high variability both within individual users across sessions and between different users, requiring robust classification approaches that can adapt to changing neural patterns while operating within computational constraints of real-time embedded systems \cite{zhang2020survey}.

Modern classification approaches for real-time intent decoding have evolved from traditional machine learning methods to sophisticated deep learning architectures. While support vector machines (SVM) and linear discriminant analysis (LDA) remain valuable for their computational efficiency and interpretability in resource-constrained applications \cite{krusienski2006toward, blankertz2011berlin}, deep learning models have demonstrated superior performance in capturing the complex, nonlinear relationships inherent in neural signals. Convolutional neural networks (CNNs) excel at extracting spatial patterns from multi-channel EEG data \cite{bashivan2015learning, manor2016convolutional}, while recurrent neural networks (RNNs) and their variants capture temporal dependencies essential for decoding dynamic motor intent. Graph neural networks (GNNs) have emerged as particularly promising for modeling the complex connectivity patterns between brain regions during motor planning and execution \cite{song2022eeg, zheng2020investigating}. These advanced architectures enable more nuanced decoding of user intent, which is vital for shared autonomy systems that must distinguish between different levels of user engagement and confidence to appropriately modulate autonomous assistance.

The integration of these technical advances allows for shared autonomy capabilities that define modern assistive BCI systems. Real-time classifiers combined with with transfer learning and adaptive filtering, allow systems to not only decode explicit user commands but also infer user confidence, attention levels, and intended goals \cite{jayaram2016transfer, kang2009composite}. This multi-layered understanding enables proactive assistance where the autonomous component can anticipate user needs, provide corrective guidance, and smoothly transition between different levels of autonomy based on decoded neural states. Such capabilities are particularly valuable in assistive and rehabilitative applications, where users may have varying levels of motor control and cognitive load, requiring systems that can dynamically adapt their assistance strategies. Future developments should focus on co-adaptive algorithms that simultaneously optimize both human neural patterns and machine learning models \cite{orsborn2014closed, danziger2009learning}, hybrid sensor integration to complement EEG with other modalities, and energy-efficient edge computing architectures to support deployment in portable, long-term use scenarios \cite{lotte2022coadaptive}.

\subsection{Human-centered performance metrics in BCI Shared Autonomy}
Evaluating autonomy in BCIs requires a multidimensional performance assessment framework that captures both system-level efficiency and user-centered outcomes. Traditional performance metrics—such as task success rate, classification accuracy, and information transfer rate (ITR)—remain widely used in BCI literature \cite{Wolpaw2002, lotte2018review}. However, these metrics alone are insufficient for evaluating the quality of shared autonomy, which involves both machine performance and human experience.
Recent studies emphasize the importance of human-centered metrics such as user workload, typically assessed through instruments like NASA Task Load Index (NASA-TLX) \cite{hart1988development}, perceived control and trust, which influence user engagement and acceptance \cite{chen2014supervisory, zhao2023human}, system adaptability and responsiveness, which reflect how well the system co-adapts to user behavior \cite{lotte2022coadaptive}, and user satisfaction or usability, often quantified using tools like the System Usability Scale (SUS) \cite{brooke1996sus}.
A major challenge lies in developing standardized benchmarking protocols that account for the dynamic interaction between user and system. For example, systems that assist more aggressively may improve task success but reduce user agency—a trade-off that must be captured through joint autonomy-performance metrics \cite{zhao2023human}.
To holistically evaluate shared autonomy, future work should consider hybrid frameworks that combine quantitative performance (e.g., accuracy, speed), behavioral indicators (e.g., error correction patterns), and qualitative feedback (e.g., user-reported metrics).

\subsection{Case Study: Graph-based AI translation for shared control in BCI grasping systems}

\begin{figure}[htbp]
    \centering
    \includegraphics[width=0.9\textwidth]{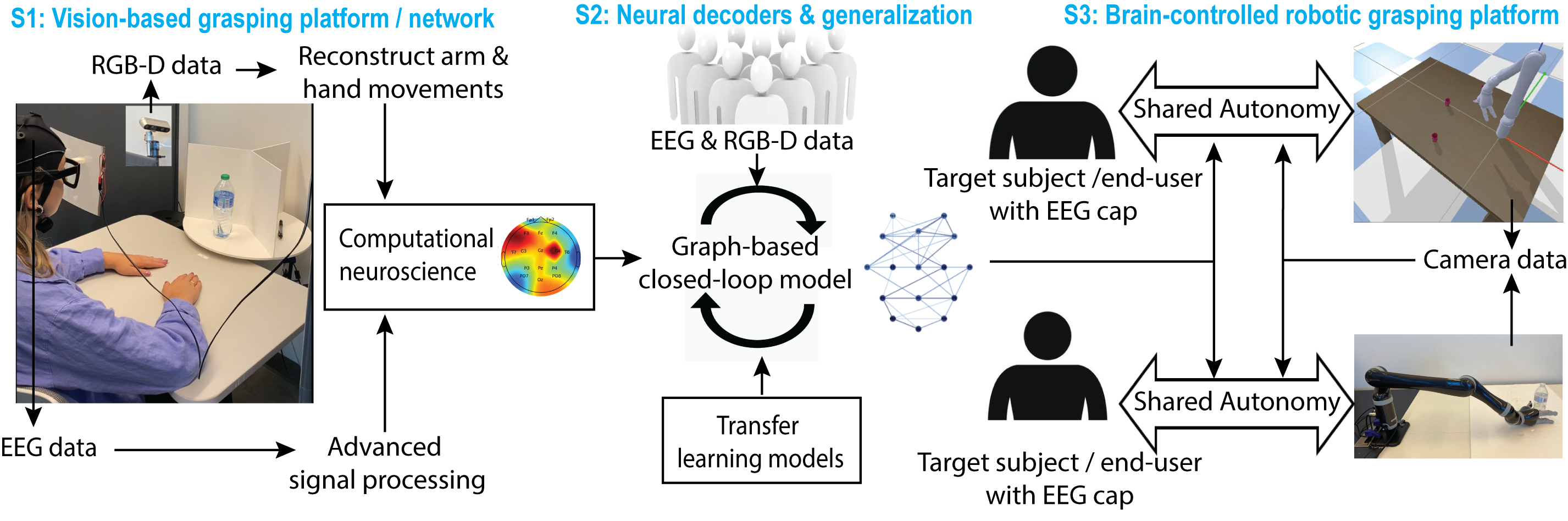}
    \caption{Graph-based AI translation pipeline for BCI grasping system with three components: S1 collects RGB-D and EEG data during reach-to-grasp movements, S2 uses graph-based neural decoders with transfer learning for motor intent classification, and S3 implements shared control between decoded neural signals and robotic autonomy via Kinova JACO arm.}
    \label{fig:bci_case_study}
\end{figure}

The role of AI as a translator of neural intent is demonstrated in our bio-inspired, multimodal BCI system for assistive grasping (Figure~\ref{fig:bci_case_study}). The pipeline begins with Stage 1 (S1), a vision-based grasping platform that collects synchronized RGB-D and EEG (tEEG) data as users observe and plan reach-to-grasp movements. This multimodal approach combines neural signals from users wearing a 16 channel EEG recording system (g.tec) with RGB-D visual information for reconstructing arm and hand kinematics through camera-based pose estimation. Advanced signal processing modules extract relevant spatial, spectral, and temporal features from the EEG data while simultaneously processing the behavioral signals captured through computer vision. These preprocessed biosignals then feed into Stage 2 (S2), where computational neuroscience approaches utilize graph-based neural decoding models that capture dynamic functional connectivity patterns across EEG channels—modeling the brain as a network to decode motor intent more accurately than traditional sequential processing methods.

In S2, we utilize transfer learning to generalize across subjects, where pretrained models learn shared neural representations related to specific grasp types and are fine-tuned for individual users with minimal data to improve usability and reduce calibration time. The decoded intent is then integrated in Stage 3 (S3), the brain-controlled robotic grasping platform, where AI continuously blends user-driven neural input with autonomous robotic feedback to control a Kinova JACO robotic arm under a shared autonomy framework. This final stage enables both target subjects and end-users to participate in smooth, adaptive interaction through the bidirectional shared autonomy paradigms shown in the figure. The co-adaptive loop—driven by real-time inference from the graph-based model and camera data integration—exemplifies the vision of bio-inspired AI, where human and machine collaborate to complete complex tasks like object grasping in a way that mimics natural sensorimotor coordination while reducing the cognitive burden inherent in traditional BCI approaches. This shared autonomy BCI framework represents a paradigm shift from direct control BCIs toward user-in-the-loop systems collaborative systems that intelligently balance user agency with autonomous assistance.

\section{Shared Autonomy in Rehabilitation Domain}
\subsection{Overview and Gaps}

Stroke is a leading cause of death and often results in long-lasting upper limb motor disabilities \cite{anwer2022rehabilitation}. Physical therapy is crucial for helping survivors regain motor control, but traditional methods and clinical systems alone can't meet the high demand due to the resource-intensive nature of therapy \cite{khalid2023robotic}. As rehabilitation shifts towards enabling patients to practice and regain strength independently, upper limb robotic rehabilitation has gained prominence \cite{khalid2023robotic}. These platforms are increasingly becoming more portable and cost-efficient, catering to various upper limb activities ranging from finger manipulation to shoulder movements \cite{qassim2020review, maciejasz2014survey, demofonti2021affordable, narayan2021development, johnson2017affordable}. Studies comparing robotic rehabilitation with traditional methods have shown positive impacts \cite{lum2002robotic, khalid2023robotic, mansour2022efficacy}. However, the widespread adoption of these devices is hindered by several factors. Complex designs often lead to higher costs, making them less affordable. Additionally, usability is a concern, especially for individuals with severe motor impairments who need simpler tasks and user-friendly interfaces. Beyond hardware considerations like cost and portability, a more significant challenge lies in refining the control strategies and adaptive algorithms that mediate the robot's real-time interaction with the patient.

\subsection{Applications and Mechanisms}
Upper limb rehabilitation robots are primarily designed as either end-effector-based or exoskeleton-based systems \cite{qassim2020review, maciejasz2014survey, demofonti2021affordable, narayan2021development}. Exoskeleton-based robots, with multiple degrees of freedom, allow for a wider range of movements due to more connection points.  In contrast, end-effector-based robots are predominantly adopted due to their simplicity, ease of production and control, and adaptability to different arm sizes of subjects. Additionally, they have the lowest cost and the greatest potential for commercialization \cite{sheng2016bilateral}.  The study \cite{lee2020comparisons} has proven the effectiveness of end-effector-based devices over exoskeleton-based robots. These end-effector robots are further categorized into three types: robots for manipulation \cite{metzger2011design, devittori2022automatic, baniasad2011wrist, buschfort2010arm}, for reaching \cite{micera2005simple, krebs2004rehabilitation, lu2011development, sivan2014home, mazzoleni2013upper, kopke2020feasibility, sulzer2007design, chang2018semi, colombo2007design, ceccarelli2021operation, wu2019adaptive, campolo2014h, freeman2009robotic, ghannadi2018configuration, guillen2021usability, whitall2000repetitive, vergaro2010self, hesse2007new, mahoney2003robotic, rashedi2009design, qian2021quantitative, avizzano2011motore, paolucci2021robotic, 10930540}, and for both manipulating and reaching \cite{5209509, rahman2015irest}. While robots designed for reaching activities are often prioritized in rehabilitation due to their focus on training upper body movements prior to hand manipulation, they also cater to a wide range of patient needs \cite{basteris2014training}. Starting with a simple design of just one degree of freedom (DOF) for those with severe conditions, these robots can be adapted to include up to three DOFs, facilitating movement training across three axes. Additionally, by primarily operating within a two-dimensional plane, these robots maintain ease of operation, implementation, and user-friendliness, further enhancing their utility across different phases of rehabilitation therapy.

There are various platforms for upper limb extremity rehabilitation. they differ from their hardware designs to their control strategies and treatment modes.  These platforms could generally be categorized into two major groups; bilateral, where both the affected and unaffected arms are involved, and unilateral, where only the affected arm participates in training \cite{van2012systematic, sheng2016bilateral}.   Additionally, the physical design of these robots can be categorized into various forms, such as H-shaped, triangular, linkage-based, and mobile, among others. Equally crucial is how these robots are programmed and controlled to adapt to the diverse requirements of different therapeutic processes. 

\subsection{Control strategies}

\begin{figure}
    \centering
    \includegraphics[width=0.9\linewidth]{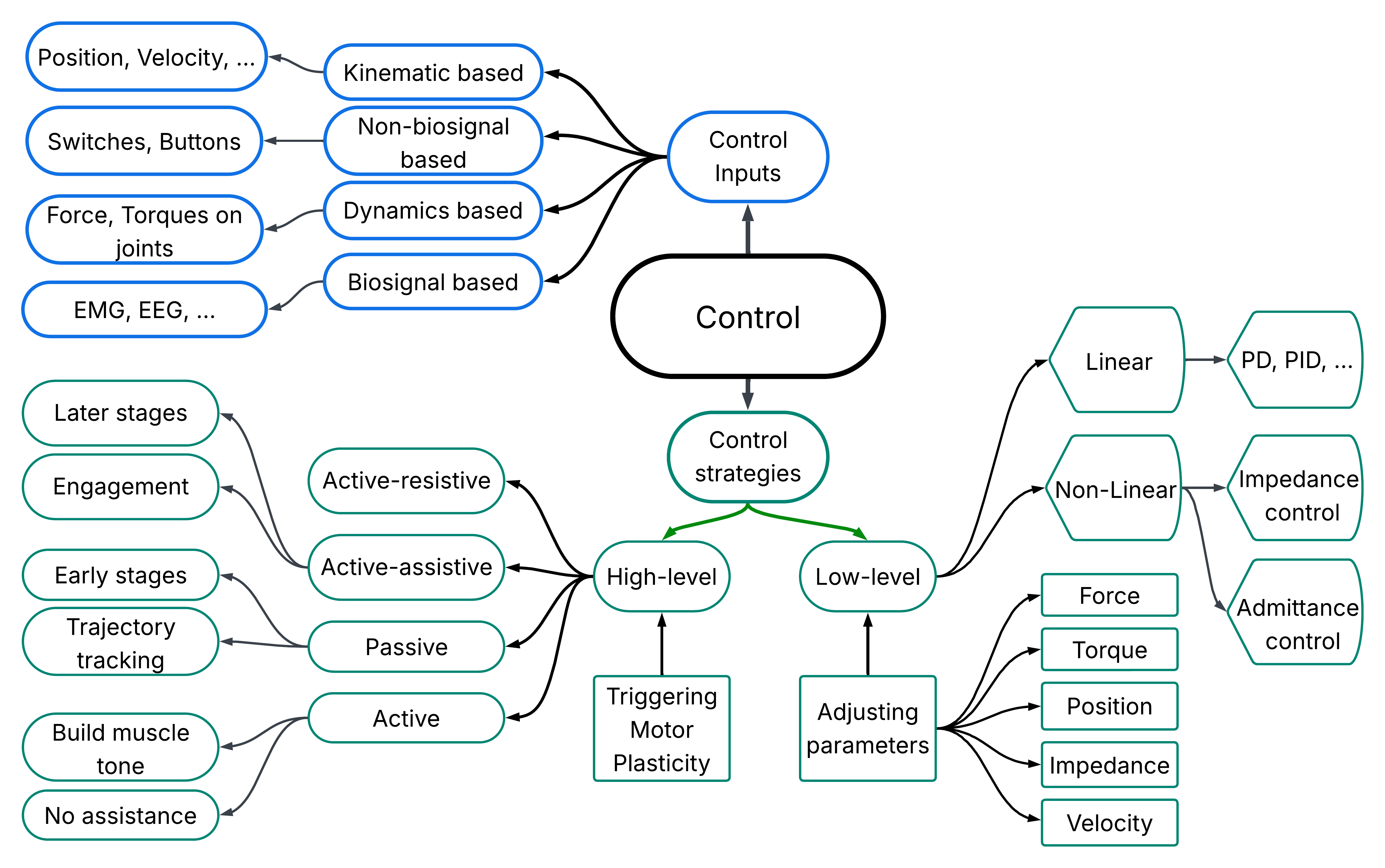}
    \caption{Hierarchical taxonomy of control inputs and strategies for upper-limb rehabilitation robots. }
    \label{control}
\end{figure}

Developing control strategies plays a crucial role in designing rehabilitation robots, particularly for assisting stroke survivors in restoring limb functions \cite{dalla2021review, proietti2016upper, basteris2014training, maciejasz2014survey, narayan2021development}. The choice of control strategy is tailored to the individual needs of the patient and the severity of their motor control impairment, forming the foundation for shared autonomy frameworks where human intent and robotic assistance must be seamlessly integrated. These strategies fall into two main categories: high-level control and low-level control (Figure \ref{control}). High-level control concentrates on triggering motor plasticity, whereas low-level control focuses on overseeing and adjusting a range of factors such as position, velocity, force or torque, impedance, admittance, and processing the adaptive feedback from high-level control. This hierarchical structure enables dynamic allocation of control authority between the patient and robot. High-level control strategies can be further classified into four main types: passive, active-assistive, active, and active-resistive, each offering different levels of patient autonomy and robot intervention.

In the early stages of rehabilitation, the passive mode is predominantly used, focusing on enhancing the range of motion and reducing spasticity in the affected limb \cite{proietti2016upper}. In this phase, the rehabilitation robot is programmed to assist patients in smoothly following a predetermined path. The initial control strategy often implemented is motion control or trajectory tracking, typically in its passive form. This method ensures that the limb follows a set trajectory within the robot’s capabilities, without necessarily requiring active effort from the patient. 

Active-assistive control strategies become prominent as therapy advances and patients start regaining control of their limbs \cite{dalla2021review}. This approach, crucial in later stages of rehabilitation, provides just enough assistance for patients to complete tasks, thus encouraging them to actively participate and exert effort in exercises. This method is tailored to promote better engagement and more effective rehabilitation outcomes, especially as patients' engagement in the exercise becomes increasingly important.

Active, or "transparent," modalities in rehabilitation robots allow patients to move without assistance or resistance from the robot, turning it into a measurement tool during assessments \cite{maciejasz2014survey}. It's crucial to maintain this transparency to avoid influencing the patient's performance. In later recovery stages, when patients need to build muscle tone, resistive modalities are used. These don't follow preset paths but instead challenge the patient with resistance proportional to their movement speed, enhancing engagement and muscle strengthening in robot-assisted exercises. Generally, these platforms primarily utilize passive and semi-active control strategies \cite{basteris2014training}, often avoiding fully active or resistive modes that are intended for patients who have largely regained their abilities. 

Low-level control groups in rehabilitation merge to support high-level objectives \cite{maciejasz2014survey, dalla2021review, narayan2021development}. These groups are mainly divided into linear and nonlinear control approaches. Linear control encompasses basic methods such as Proportional-Derivative (PD), Proportional-Integral-Derivative (PID), and their variants. However, due to the nonlinearity in the dynamics of upper extremity exoskeletons, arising from flexible links, joints, and actuators, nonlinear control methods are preferred to enhance system robustness and compatibility with external factors. Nonlinear control includes techniques like computed torque control, impedance control, sliding mode control, adaptive control, and admittance control.

\begin{figure}[htbp]
  \centering
  \includegraphics[width=0.8\textwidth]{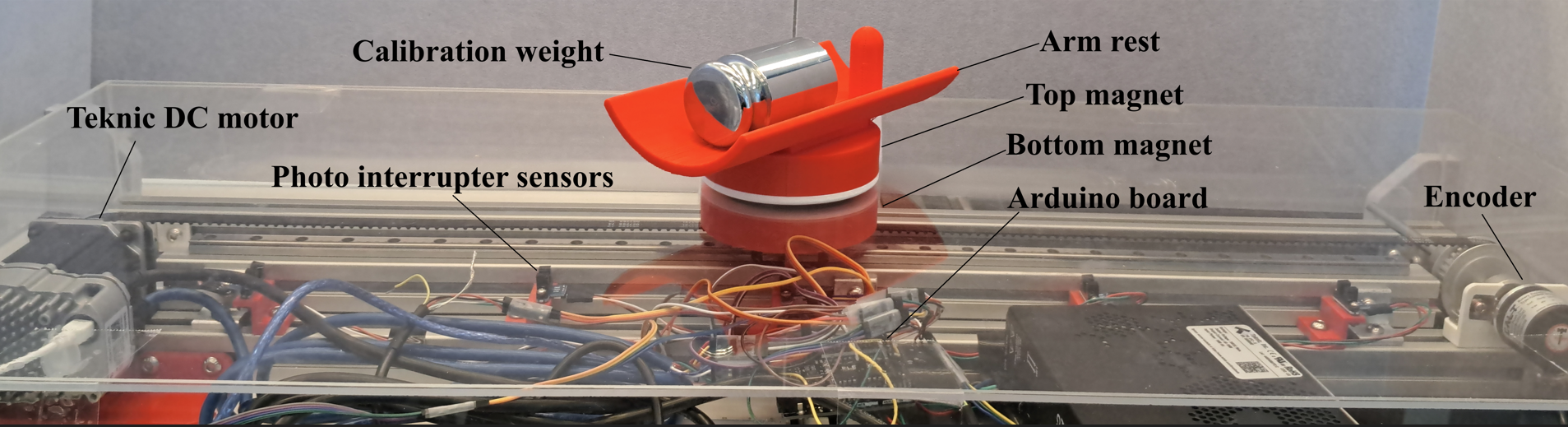}
  \\[0.3cm] 
  \includegraphics[width=0.8\textwidth]{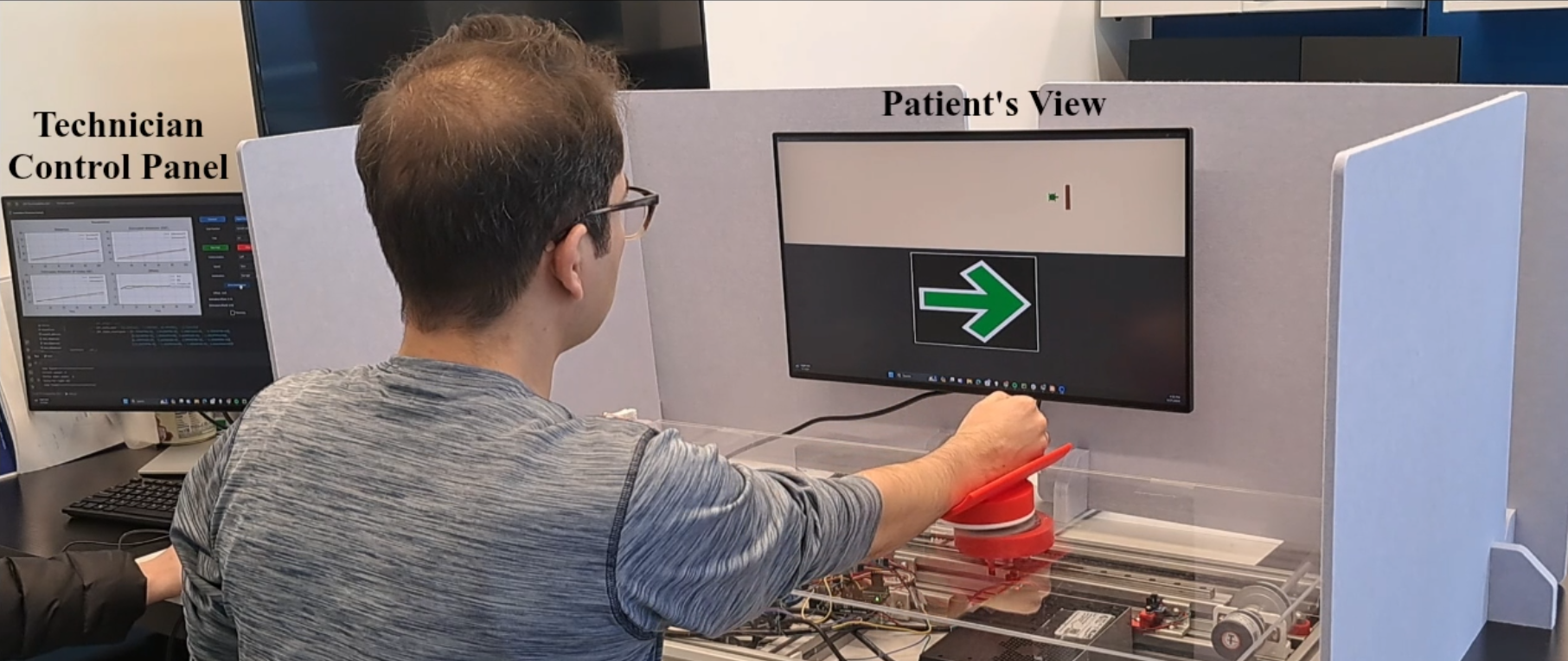}
  \caption{Experimental setup demonstrating user interaction with the magnetic-based robotic rehabilitation platform. The designed planar robotic platform (top), a preview of running a trial with a participant. (bottom)}
  \label{planar}
\end{figure}

\subsection{Biosignal Integration for Shared Control}

Control inputs are a vital element in defining the control architecture of rehabilitation devices, and they are divided into four main categories \cite{narayan2021development, desplenter2020rehabilitative} (Figure \ref{control}). Kinematic-based inputs record orientations, positions, velocities, and accelerations of specific parts of the device or limb. Dynamic-based inputs focus on the force or torque applied by the user on the joints or ends of rehabilitation devices. Biological-based inputs use EMG (and sometimes EEG) signals in an on/off or proportional fashion to align device movements with the user’s motor intent. Finally, non-biological-based inputs involve switches or buttons to detect human intentions and initiate limb movements. While kinematic and dynamic sensors remain the most common in clinical platforms, integrating biosignals enables a richer, more intuitive human–robot interaction.

\subsection{Adaptive algorithms} Building on the control strategies and input modalities described above, adaptive algorithms serve as the critical mechanism that customizes robotic assistance to each patient’s real-time needs. By integrating kinematic data (e.g., joint angles and velocities), dynamic feedback (interaction forces or torques), and biosignals (such as EMG or EEG indicators of muscle engagement), these algorithms continuously recalibrate low-level control loops—whether impedance, admittance, or direct torque control—to deliver the appropriate high-level mode of assistance (passive, active-assistive, active, or active-resistive). In model-reference adaptive schemes, the controller continuously compares the actual limb trajectory to a predefined reference model, adjusting control gains in real time to accommodate uncertainties in patient dynamics and human–robot interaction. For repetitive tasks, an iterative learning component retains tracking errors from previous trials and refines feedforward commands to promote faster convergence toward smooth, accurate movements. Adaptive impedance and admittance controllers modulate virtual stiffness and damping based on real-time assessments of muscle effort—providing minimal resistance during passive or assistive phases and gradually increasing challenge as voluntary engagement intensifies. More advanced neuro-adaptive and fuzzy-logic approaches utilize neural networks or fuzzy estimators to model complex, nonlinear patient–robot interactions without relying on explicit system representations. These techniques enable progressive refinement of control strategies across therapy sessions, facilitating personalized interventions that adapt to each patient’s evolving motor capabilities and promote more effective rehabilitation \cite{narayan2021development, proietti2016upper}.

\subsection{Case study: Stroke rehabilitation framework}
The work from \cite{10930540} presents a novel robotic rehabilitation framework designed specifically for passive stroke rehabilitation using a one-dimensional magnetic-based actuation mechanism (Figure \ref{planar}). The system targets transitional shoulder movements (flexion-extension) and employs a non-contact magnetic interface to ensure smooth, safe, and user-friendly interaction. Replacing conventional mechanical linkages, the platform minimizes physical strain and risk while enhancing motion quality and weight compensation. The setup includes a magnetically actuated end-effector and an ergonomic armrest, with motion tracked and stabilized via an Extended Kalman Filter (EKF) for robust control, even under sensor failure. In trials with 12 participants, users reported high satisfaction in terms of safety, comfort, and system responsiveness. The platform demonstrated stable, synchronized motion and low tracking error, validating its potential for home-based early rehabilitation. While this initial design supports passive therapy along a single linear axis, future development aims to expand to two-dimensional movements and incorporate assistive control strategies, enabling more active participation and broader functional training for stroke survivors.

\section{Shared Autonomy in Assistive Technologies Domain}

\subsection{Overview and gaps}

Assistive technologies encompass a broad range of domains aimed at supporting individuals with disabilities in daily activities. These systems can be categorized into mobile servant robots, physical assistant robots, wearable suits/sensors, and person carrier robots. \cite{international2014robots}. In practice, this translates to devices such as wheelchair-mounted robotic manipulators (servant robots) \cite{jiang2016enhanced}, wearable exoskeletons/prosthetics/sensors  \cite{chapman2025accuracy, de2016exoskeletons}, and intelligent powered wheelchairs (person carriers) \cite{cooper1995intelligent}. Each of these assistive devices addresses different needs – for example, a robotic arm can help a person with quadriplegia grasp and drink from a cup independently \cite{rabiee2025learning}, whereas a smart wheelchair with obstacle avoidance can enhance mobility for those who cannot safely drive a standard power chair on their own \cite{udupa2023shared}. A representative focus has long been on wheelchair-mounted assistive robotic manipulators \cite{rabiee2024streams, park2020active, abiri2024toward}, which give users the reach and dexterity needed for ADLs such as feeding, grooming, and object retrieval. The Kinova Jaco arm remains the most widely deployed example; its 6-DoF carbon-fiber structure and three-finger gripper are now reimbursable by U.S. state Medicaid programs, reflecting growing clinical acceptance \cite{ackerman2019jaco}. Other commercial devices include the Dutch iARM \cite{blom2017assistive}, descended from the original MANUS project and designed for intuitive joystick or app control, and the U.S.-developed DEKA “LUKE” prosthetic arm \cite{guizzo2014dean}, the first multi-articulated upper-limb prosthesis to receive FDA clearance, offering up to ten powered DoF for amputees. 

Despite the availability of such assistive systems, because the control channels available to people with motor impairments are inherently low-bandwidth, noisy, and fatigue-inducing, asking users to tele-operate multi-DoF arms, exoskeletons, or power chairs in a purely manual fashion is often impractical and slow \cite{jain2015assistive}. Conversely, turning the devices into fully autonomous robots can leave users feeling disempowered and vulnerable when the autonomy misinterprets goals or environmental nuances \cite{dragan2013policy}. Shared-autonomy paradigms integrate intelligent assistance directly into these devices to bridge this gap: the human provides strategic intent and retains final authority, while the onboard autonomy supplies continuous low-level stabilization, obstacle avoidance, and motion refinement. Empirical studies consistently show that this collaborative division of labor lowers task time and error rates, preserves the user’s sense of agency, and increases overall satisfaction compared with either manual or fully automated control schemes \cite{jain2015assistive, dragan2013policy}. For these reasons, shared control has become a foundational design principle across assistive domains, ensuring that advanced robotics enhances, rather than supplants, human capability.

\subsection{Shared Autonomy Architectures}

\begin{figure}[ht!]            
  \centering                    
  \includegraphics[width=0.6\linewidth]{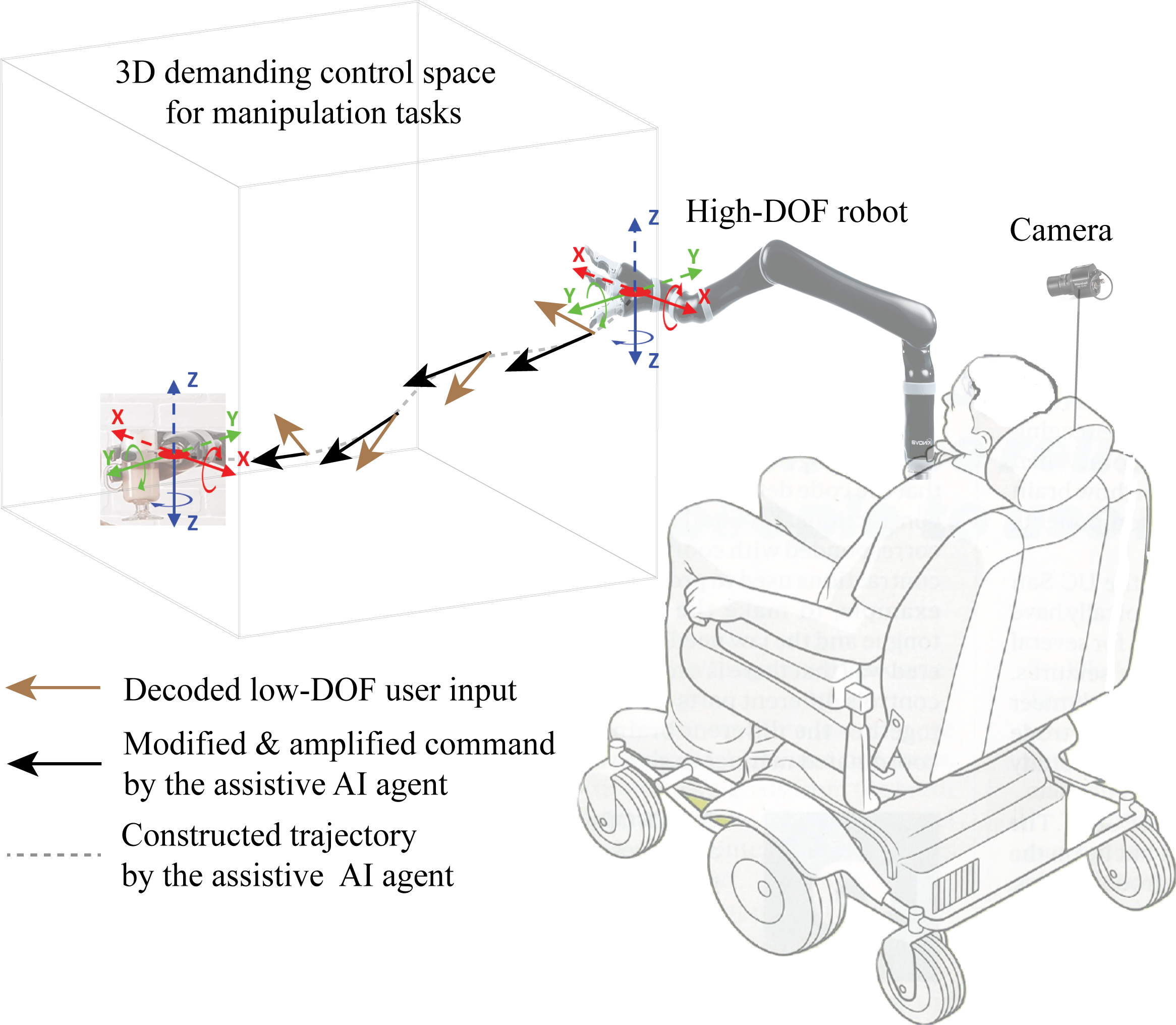}
  \caption{A conceptual design of an assistive AI-powered robotic manipulation system amplified low-dimensional user inputs for high-DoF tasks.}
  \label{fig:conceptual_img}         
\end{figure}

In the context of assistive technologies, shared autonomy (also called shared control) refers to control approaches that combine the user’s input with the robot’s autonomous actions to achieve the user’s intended goals \cite{atan2024assistive}. In these systems, the user typically provides high-level commands or coarse guidance, and the autonomy module refines and executes low-level actions, effectively amplifying the user’s capabilities \cite{rabiee2025learning} (Fig.  \ref{fig:conceptual_img}). Early traditional shared control systems often relied on manual mode switching \cite{herlant2016assistive, rudigkeit2014towards}, requiring the user to sequentially control different degrees of freedom (e.g., selecting “move arm base” vs “rotate wrist” modes) to cope with limited input dimensionality. However, frequent mode switches drastically slow down task execution and impose a high cognitive load on the user \cite{surale2017experimental}.

To relieve that burden, researchers introduced task-aware architectures that embed contextual knowledge of the environment and of likely user goals \cite{gao2014contextual, abiri2024toward}. These architectures use contextual and goal information to assist the user accordingly, rather than blindly executing pre-programmed motions \cite{gopinath2016human}. For example, if a user is teleoperating a robot arm to pick up a cup, a task-aware system will incorporate knowledge of the cup’s location and the goal of grasping when deciding how to assist. A central element of these architectures is the arbitration mechanism, the policy that decides how to blend or switch between user control and autonomous control at each moment \cite{oh2019learning}. Designing this arbitration is non-trivial: too little autonomy yields high user workload, while too much can erode the user’s sense of control \cite{shneiderman2020human}. Modern approaches, therefore, strive for user-centric, adaptive arbitration, adjusting the level of assistance based on the context and the individual user’s needs \cite{reddy2018shared}. Many frameworks formalize this as a partially observable decision problem \cite{yow2023shared, javdani2018shared}: the system maintains a belief over possible user goals and selects autonomous actions that best advance the likely goal. When the user’s goal is known or explicitly specified, assistance can be planned via standard Markov Decision Processes (MDPs) for the task. If the goal is uncertain, algorithms based on Partially Observable MDPs (POMDPs) or Bayesian inference are employed to infer intent and plan accordingly \cite{dragan2013policy}. This approach was demonstrated in prior works \cite{shafti2019gaze} that predict the user’s intended target from their initial joystick motions or gaze, then autonomously guide the robot toward that target. The result is an intelligent assistive policy that, for instance, can start moving the arm toward a likely object of interest even before the user provides a perfectly precise command \cite{downey2016blending}.

Building on these task-aware arbitration frameworks, recent research has begun to replace hand-engineered heuristics with end-to-end learning components, often deep reinforcement learning (RL) agents \cite{rabiee2024streams, rabiee2025learning, reddy2018shared}, trained in high-fidelity simulation, that can discover when to yield, when to assist, and how to interpolate smoothly between those extremes. By optimizing over millions of synthetic episodes, such agents learn to embed intent inference, motion refinement, and obstacle avoidance inside a single policy, eliminating the need for explicit mode switches or manually tuned sub-modules.

\subsection{User-Intent Inference \& Personalization Pipeline}

At the heart of shared autonomy lies the challenge of inferring the user’s intent from limited and noisy inputs, and then personalizing the assistance to the user’s abilities and preferences \cite{jain2019probabilistic}. Users with severe motor impairments may control assistive robots through a variety of interfaces: examples include low-dimensional inputs like a two-axis joystick, sip-and-puff switches, head tilts, eye-gaze trackers, or brain-computer interface signals (EEG/EMG) \cite{xu2023transitioning, beraldo2022shared, rabiee2024comparative, 10782674, ghafoori2024bispectrum, cetera2024classification}. These input channels typically provide far fewer degrees of control than the robot has degrees of freedom – for instance, controlling a 7-DoF arm with a 2-DoF joystick – and the signals can be noisy or intermittent. Therefore, the system must decode the user’s intended action or goal from whatever cues it can observe. In non-invasive setups, this inference is complicated by artifacts and user fatigue: surface EEG/EMG signals require extensive user training to be reliable, and even then, they carry limited information bandwidth \cite{tai2008review, Wolpaw2002}. Similarly, a user who can only issue simple binary commands (e.g. a head switch for left/right) cannot directly teleoperate a robotic arm to reach and grasp an object without intelligent mediation \cite{mugler2010design}.

To tackle this, assistive shared autonomy systems employ a pipeline of intent inference algorithms. One common approach is goal prediction: as the user begins an action, the system compares the motion or commands to a set of possible targets or tasks \cite{stepputtis2020language, shafti2019gaze}. For example, if a robotic wheelchair sees several doorways and the user’s initial joystick push is roughly toward one of them, a Bayesian estimator or classifier can infer a high probability that the user intends to go through that doorway. In robotic manipulators, techniques like inverse reinforcement learning \cite{ziebart2009planning} and Bayesian intent prediction \cite{best2015bayesian} have been used to guess which object a user is reaching for, after observing a portion of the motion trajectory. Notably, when no explicit goal selection is provided, the system can maintain a probability distribution over possible goals and continuously update it as more user input comes in. Javdani et al. \cite{javdani2018shared} implemented this via a POMDP-based shared autonomy framework where the human’s control input is treated as observations about the goal, and the robot’s autonomous policy aims to maximize expected progress towards the inferred goal. This allows the robot to assist proactively, for example, moving toward the most likely target while still being ready to switch if the user’s input indicates a different intention.

Beyond goal inference, the personalization aspect of the pipeline adapts the system’s behavior to individual users. Users have varying reaction times, precision, strength, and preferences for how much help they want. An adaptive shared-control system can learn these characteristics over time. One strategy is to adjust the assistance level (autonomy gain) based on performance \cite{oh2019learning}: if the user is struggling (e.g., taking too long or issuing erratic commands), the system can increase autonomy; if the user seems adept, the system can relinquish more control. Atan et al. \cite{atan2024assistive} introduced an arbitration method that monitors the user’s task performance and expertise level, and dynamically tunes the autonomy assistance at runtime. This user-centric approach led to higher efficiency in a reaching and grasping task by giving novices more assistance and experts more manual control, aligning with each user’s abilities. Another facet of personalization is intent interpretation tuning: the system may learn to interpret a particular user’s input idiosyncrasies. For example, one person’s neck movements might consistently overshoot targets, so the system’s mapping of head tilt to cursor movement can be customized (through calibration or learning) to be less sensitive. Similarly, a shared autonomy framework can incorporate a model of the user’s reaction or comfort. Some recent work proposes using reinforcement learning with human feedback to adjust how aggressively or conservatively the robot assists, based on the user’s satisfaction signals \cite{reddy2018shared}.

\subsection{Evaluation, Validation \& Regulatory Path}

Evaluating a shared-autonomy device starts with technical metrics such as success rate, completion time, motion precision, and the reduction in user effort.  Studies of wheelchair-mounted arms, for example, routinely compare how quickly and reliably users can grasp objects under direct control versus shared control \cite{rabiee2025learning}.  Equally important are subjective workload and agency measures: tools such as the NASA-TLX questionnaire quantify perceived cognitive load in human-robot collaboration\cite{javernik2023nasa}, while recent sense of agency experiments show a predictable trade off, greater robot autonomy boosts task performance but can diminish the operator’s feeling of control \cite{collier2025sense}.  Hence, optimal systems aim for adaptive, mid-spectrum autonomy that lifts efficiency yet preserves user empowerment.

Validation then progresses to end-user trials, safety certification, and long-term reliability checks.  Because recruiting clinical populations can be difficult, developers often prototype in simulation frameworks such as Assistive Gym, which model ADL scenarios and human biomechanics before moving to real users \cite{erickson2020assistive}.  When devices reach the clinic or home, they must satisfy standards like ISO 13482 for personal-care robots, demonstrating compliant forces, failsafe behaviors, and predictable fault handling \cite{international2014robots}.  Cybersecurity, data privacy, and wear-and-tear robustness add further regulatory hurdles, which partly explains why many laboratory prototypes never reach the market.  Nevertheless, evolving testbeds and clearer standards are lowering these barriers, fostering a new generation of assistive robots that are not only capable but also trusted for everyday use.

\subsection{Case Study: Adaptive Reinforcement learning for Amplification of limited inputs in Shared autonomy (ARAS)}
\begin{figure}[ht!]            
  \centering                    
  \includegraphics[width=0.6\linewidth]{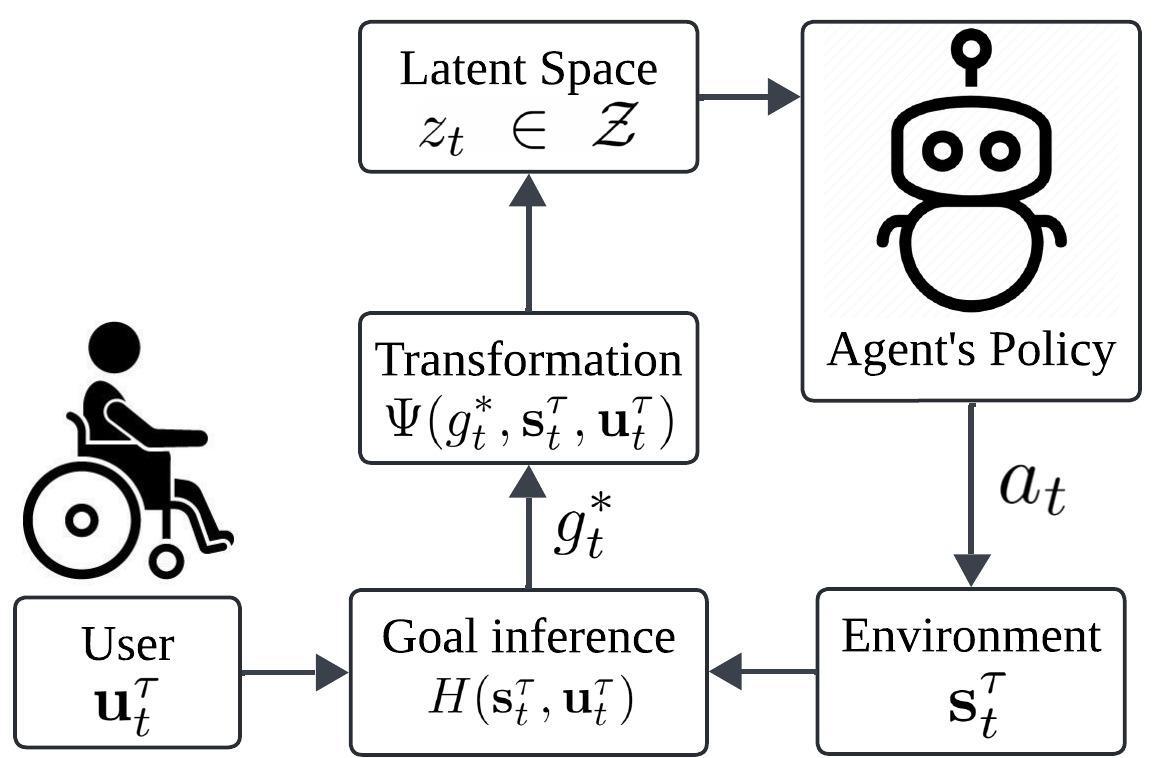}
  \caption{ARAS formulation diagram.}
  \label{fig:ARAS_framework}         
\end{figure}

As a concrete illustration, Rabiee et al. \cite{rabiee2025learning} introduced the Adaptive Reinforcement learning for Amplification of limited inputs in Shared autonomy (ARAS) framework, whose high-level dataflow is sketched in Fig. \ref{fig:ARAS_framework}. At each control cycle, the system receives a short window of low-bandwidth user commands $u^{\tau}_{t}$ together with scene information $s^{\tau}_{t}$ from an overhead RGB camera. A goal inference module $H(s^{\tau}_{t},u^{\tau}_{t})$ maintains a Bayesian belief over the user’s intention and returns the most likely goal $g^{*}_{t}$. These signals are fused by a transformation $\Psi(g^{*}_{t},s^{\tau}_{t},u^{\tau}_{t})$ that filters irrelevant pixels and embeddings, producing a compact latent state $z_{t}$ that captures only task-relevant features. A deep Q-network then maps $z_{t}$ to a high-level robot action $a_{t}$, which the physical arm executes; the resulting state feeds back into the loop, allowing the policy to amplify limited inputs into smooth 7-DoF manipulation without mode switching or explicit mid task queries. In essence, ARAS lets the human supply strategic intent while the autonomy layer handles continuous motion planning and goal refinement in real time.

Extensive evaluation confirms the benefits of this formulation. After 50,000 simulated pick-and-place episodes, the learned policy transferred zero-shot to hardware, where 23 participants completed 90 trials apiece. Compared with manual teleoperation, ARAS cut average completion time, reduced required input keystrokes, and lifted overall success from 86\% to 92.8\ while preserving user agency. In simulation, ARAS maintained near-perfect performance (99.8\% success, 31 s completion) under fixed goals and still exceeded 96\% success when users changed intentions midtask, outperforming both a vanilla DQN and a hindsight-optimization baseline across all scenarios. Subjective NASA-TLX scores and Likert surveys showed significant drops in perceived workload and the highest ratings for satisfaction, flexibility, and future use likelihood, indicating that users valued the system’s blend of help and control. Together, these results demonstrate that a single latent-space RL policy can infer, amplify, and adapt low-DoF inputs at interactive rates, delivering high-precision shared autonomy that meets both technical and experiential benchmarks.

\section{Emerging Trends}

\subsection{Large language Models in Human-centered Shared Autonomy}

Large-language-model (LLM) techniques are rapidly reshaping shared autonomy by turning natural language into a control modality and by supplying robots with onboard reasoning. This development allows LLMs to function as adaptive, dialogue-native co-pilots rather than post-hoc interfaces, due to advancements in their reasoning capabilities and multimodal interfaces.

\subsubsection{Reasoning Capabilities}

Large Language Models (LLMs) are evolving from simple command-followers into reasoning engines for shared autonomy, enabling them to handle ambiguous goals, plan complex tasks, and apply commonsense knowledge. Early demonstrations such as SayCan \cite{ahn2022can} grounded a frozen LLM’s high-level plans in low-level affordance scores, allowing a mobile manipulator to execute multistep kitchen tasks from free-form instructions with a 74\% real-world success rate (double that of a language-only baseline). More recent models push this idea further; PaLM-E \cite{driess2023palm} interleaves text tokens with raw images and proprioceptive states inside a single transformer, allowing one network to generalize across navigation, grasping, and visual question-answering. This reasoning process can be made more robust through closed-loop deliberation. For instance, the Inner Monologue framework \cite{huang2022inner} keeps the LLM “thinking out loud,” injecting real-time success detection and scene descriptions back into its context so the robot can re-plan or query the user if execution goes off course.

Beyond direct execution, LLMs are also used as synthesis engines to make robust policies. Code-generation approaches such as RobotGPT \cite{jin2024robotgpt} treat multiple drafts from ChatGPT as demonstrations, then distill them into a reinforcement-learning policy that boosts manipulation success from 38\% to 91\% on pick-and-place tasks. A key aspect of advanced reasoning is also understanding the model's own limitations. To address the failure modes and uncertainty of LLM planners, the KNOWNO framework \cite{ren2023robots} uses conformal prediction to calibrate an LLM’s confidence. When multiple high-probability next steps exist, the robot “knows it doesn’t know” and explicitly asks the human for guidance, cutting unnecessary interventions by up to 24\% while preserving task-completion guarantees.

This capacity for nuanced reasoning is being applied to decompose complex task sequences. Some frameworks combine LLMs with classical planners to guarantee logical soundness \cite{liu2024llmclassical}, while others leverage external knowledge. The AdaptBot framework, for instance, integrates an LLM with a Knowledge Graph to break down generic commands into specific actions, using human-in-the-loop dialogue to resolve ambiguities \cite{kabra2024adaptbot}. More recent work uses LLMs for object-level planning, where the model reasons about object state transitions to generate sequences of subgoals for traditional task and motion planners \cite{paul2025objectlevel}. This approach is enhanced by the LLM's embedded commonsense knowledge, which proves critical for inferring unstated user needs. The Guide-LLM framework \cite{chen2025guidellm} demonstrated this by planning safe indoor navigation paths for patients with visual impairments, inferring potential hazards using text-based maps. Similarly, the SAVOR framework \cite{wu2025savor} uses a VLM's commonsense priors about food properties (e.g., softness, shape) to select appropriate utensils and actions for robot-assisted feeding while refining its plan through visual and haptic feedback.

\subsubsection{Multimodal Interfaces}

Advanced reasoning must be paired with interfaces that support effective communication. Therefore, parallel progress in multimodal interfaces is critical, focusing on deeply integrating vision and language, generating expressive and explainable feedback, and exploring novel fusion with human biosignals to create a more seamless human-robot partnership.

Recent Vision-Language-Action (VLA) models are moving beyond single-step commands to enable long-horizon, visually-grounded tasks. Foundational models like RT-2 \cite{brohan2023rt2} have demonstrated that VLM-based models can be finetuned with robotic data to acquire secondary robotic capabilities, directly translating visual and language commands into actions. The Vision-Language Model Predictive Control framework \cite{cui2024vlmpc} exemplifies the integration of these models into the robot's control loop, using a VLM's understanding to continuously sample and predict the outcome of potential action sequences. To scale the training of such systems, the field is also developing large-scale, primitive-level datasets that break down complex tasks into a vocabulary of fundamental skills to create more robust agents \cite{fu2024rh20tp}.

To foster user trust and facilitate effective collaboration, the system must also be able to explain its reasoning. Recent LLMs are capable of generating context-aware, natural language explanations \cite{hafner2024safe, lai2023lami, tan2023reflect}. Some systems can now generate contrastive explanations, clarifying not only why a certain action was taken, but why an alternative was not, which is crucial for debugging and building accurate end-user mental models \cite{ehsani2025contrastive}. Beyond text, by combining a VLM with Grad-CAM heatmaps, a robot can also visually highlight the part of the scene that influenced its navigation decision while verbally explaining its reasoning \cite{singh2025explainable}.

\subsection{Ethical considerations and human factors}

The practical success of shared autonomy hinges on more than just technical execution and an integration of ethical and human factors is required to mitigate the risk of severe, real-world consequences. These risks extend beyond system abandonment \cite{Hayes2013} to include physical harm, psychological damage, and the reinforcement of societal inequities. The black box nature of many AI models creates challenges for transparency \cite{Cummings2004Accountability}, but it also may foster misaligned trust. Users can develop automation complacency, where they over-trust a system that is mostly reliable and become blind to the edge cases where it can critically fail, leading to physical harm \cite{Parasuraman1997}. The risk is not simply that the AI fails, but that the system's design failed to foster calibrated trust, leaving the user unaware of the system's operational boundaries.

Furthermore, a poorly designed system can even cause psychological harm. Systems that are overly assertive or that frequently override user commands can erode the user's sense of agency \cite{collier2025sense}, leading to a state of learned helplessness, which is against the the goals of rehabilitation systems \cite{Seligman1972, Maier2016}. This can cause a degradation of the user's own skills(also known as skill atrophy), which is a crucial concern in therapeutic applications \cite{Carr2019}. The user's experience can also be degraded by the "frustration of the excluded middle", where the system handles all simple tasks but disengages during complex scenarios, leaving the user with high cognitive burden precisely when they need the most support \cite{Bainbridge1983}.

Finally, unexamined systems risk amplifying societal biases. Algorithmic bias \cite{FoschVillaronga2021} can lead to discriminatory outcomes when intent recognition or computer vision models, trained on data from dominant demographic, economic, or cultural groups, fail to perform reliably for minority users \cite{Howard2019}. This can render a device ineffective or unsafe, creating scenarios where the benefits of the system are only available to a selective group, thereby limiting the systems accessibility. Addressing these factors is a pragmatic necessity for creating safe, impartial, and effective shared autonomy systems.

\section{Summary}
This chapter has provided a comprehensive analysis of human-centered shared autonomy, arguing for a common technical framework to connect the distinct applications of Brain-Computer Interfaces (BCIs), rehabilitation, and assistive robotics. The proposed framework is grounded in a core challenge: mapping noisy user biosignals to effective control policies through adaptive arbitration. Our review detailed the foundational principles for such systems, including maintaining the human as the leader in the human-AI team, implementing varied interaction paradigms, and the critical role of robust intent inference to translate user goals into action.

Across the three domains, we examined how these principles are applied to address specific needs, such as mitigating the cognitive load of direct BCI control , enabling personalized, adaptive therapy in rehabilitation , and amplifying low-bandwidth commands for high-dimensional assistive tasks. The case studies presented serve as examples of these concepts in practice. Throughout this analysis, we identified shared challenges, including the high variability of biosignals , the difficulty of robustly decoding intent , and the fundamental trade-off between task performance and the user's sense of agency.

Finally, we explored emerging trends that will shape the future of this field. While Large Language Models show promise for enabling more complex reasoning and multimodal interaction, their integration, like all advancements in this area, must be navigated with careful attention to ethical and human factors. As the paper highlights, issues of automation complacency, skill atrophy, and algorithmic bias are critical safety and usability concerns that must be proactively addressed. The trajectory of this research is toward more deeply integrated human-AI joint cognitive systems, but achieving this goal requires a steadfast focus on developing technologies that are not merely capable, but also transparent, trustworthy, and empowering for the end-user.

\bibliographystyle{IEEEtran}
\bibliography{references}

\end{document}